\documentclass[prb, showpacs,showkeys,preprint]{revtex4}

\usepackage{graphicx}

\begin{document}

\title{On the equivalence of the microcanonical and the canonical ensembles:
a geometrical approach}

\author{Ricardo L\'opez-Ruiz}
\email{rilopez@unizar.es}
\affiliation{
DIIS and BIFI, Facultad de Ciencias, \\
Universidad de Zaragoza, E-50009 Zaragoza, Spain}

\author{Jaime Sa\~nudo }
\email{jsr@unex.es}
\affiliation{
Departamento de F\'isica, Facultad de Ciencias, \\
Universidad de Extremadura, E-06071 Badajoz, Spain}

\author{Xavier Calbet}
\email{xcalbet@googlemail.es}
\affiliation{
Instituto de Astrof{\'\i}sica de Canarias, \\
V{\'\i}a L\'actea, s/n, 
E-38200 La Laguna, Tenerife, Spain}

\date{\today}

\begin{abstract}
In this paper, we consider the volume enclosed by the microcanonical ensemble in phase space
as a statistical ensemble. This can be interpreted as an intermediate image between
the microcanonical and the canonical pictures.
By maintaining the ergodic hypothesis over this ensemble, that is,
the equiprobability of all its accessible states, the equivalence of this ensemble 
in the thermodynamic limit  with the microcanonical and the canonical ensembles 
is suggested by means of geometrical arguments.
The Maxwellian and the Boltzmann-Gibbs distributions are
obtained from this formalism. 
In the appendix, the derivation of the Boltzmann factor from a new {\it microcanonical 
image} of the canonical ensemble is also given.  
\end{abstract}


\maketitle

The microcanonical and the canonical ensembles represent two clearly 
different physical situations in a statistical system \cite{huang}. 
The microcanonical ensemble is presented in the literature 
as modeling an isolated system that conserves its 
energy in time. The canonical ensemble models a system in contact with 
a heat reservoir containing an infinite energy,
which allows to fluctuate the energy of the system but maintains its mean value  
constant in time. In the thermodynamic limit, and under certain assumptions on the entropy
function \cite{ellis}, both formalisms converge and they give the same
macroscopic statistical results \cite{huang,ellis}.

Here, we interpret the volume enclosed by the microcanonical ensemble
in phase space as a statistical ensemble.
It implies the existence of some kind of heat reservoir with an upper limit
energy in order that the system can visit all the accessible states
enclosed in that volume.
The geometrical reason why this picture is equivalent to the
microcanonical ensemble in the thermodynamic limit is discussed. 
Thus, if the microcanonical ensemble is supposed to be 
represented by the equiprobability over the
hypersurface on which the system evolves as consequence of conserving an energy $E$
(as recently explained in Refs. \onlinecite{lopez2007-1,lopez2007-2}), 
then the volume-based ensemble can be interpreted  
as the equiprobability over the whole volume which is enclosed by that hypersurface.
This means that, in the latter ensemble, the system can visit states 
with different energies, with an upper limit $E$
given by the energy defined in its equivalent microcanonical picture.
As we have said, we can think that in this image the system is exchanging energy with 
a heat (or energy) reservoir containing a maximum energy $E$. 
This constraint is removed in the thermodynamic limit when the number 
of degrees of freedom and the energy $E$ are 
supposed to become infinite.  
Let us observe that this infinite limit establishes also the equivalence
between this ensemble and the canonical ensemble (when certain conditions
of smoothness in the entropy function are implicit \cite{ellis}), just because in this case
the reservoir contains an infinite energy and then both pictures become identical.  
Hence, when the number of dimensions of the system increases infinitely, 
almost all the volume enclosed by that hypersurface is located in the vanishingly 
thin layer close to the hypersurface, and, 
in consequence, surface and volume, tend to coincide.
This is the reason why the microcanonical ensemble and the volume-based
ensemble, and by extension the canonical ensemble, 
give the same results for systems well-behaved \cite{ellis} in the thermodynamic limit.

We proceed to obtain different classical results from this 
volume-based statistical ensemble. 
We start by deriving (recalling) the Maxwellian (Gaussian) distribution
from geometrical arguments over the volume of an $N$-sphere.
Following the same insight, we also explain the origin of the Boltzmann-Gibbs (exponential) 
distribution by means of the geometrical properties of the 
volume of an $N$-dimensional pyramid.
We finish claiming a possible general statistical result that follows from 
the properties of the volume enclosed by a one-parameter dependent family 
of hypersurfaces, 
in which $N$-spheres and $N$-dimensional pyramids are included. 
In the appendix, an alternative microcanonical image of the 
canonical ensemble is also given.

\section*{Derivation of the Maxwellian distribution}

Let us suppose a one-dimensional ideal gas of $N$ non-identical 
classical particles with masses $m_i$, with $i=1,\ldots,N$, and total 
maximum energy $E$. If particle
$i$ has a momentum $m_iv_i$, we define a kinetic energy:
\begin{equation}
K \equiv p_i^2 \equiv {1 \over 2}{ m_iv_i^2},
\label{eq-p_i}
\end{equation} 
where $p_i$ is the square root of the kinetic energy. 
If the total maximum energy is defined as $E \equiv R^2$, we have 
\begin{equation}
p_1^2+p_2^2+\cdots +p_{N-1}^2+p_N^2 \leq R^2.
\label{eq-E}
\end{equation} 
We see that the system has accessible states with different energy, which is 
supplied by the heat reservoir. These states are all those enclosed into the volume 
of the $N$-sphere given by Eq. (\ref{eq-E}). 
The formula for the volume $V_N(R)$
of an $N$-sphere of radius $R$ is
\begin{equation}
V_N(R) = {\pi^{N\over 2}\over \Gamma({N\over 2}+1)}R^{N},
\label{eq-S_n}
\end{equation}
where $\Gamma(\cdot)$ is the gamma function. If we suppose that each point
into the $N$-sphere is equiprobable, then the probability $f(p_i)dp_i$ of finding 
the particle $i$ with coordinate $p_i$ (energy $p_i^2$) is proportional to the 
volume formed by all the points on the $N$-sphere having the $i$th-coordinate 
equal to $p_i$. 
Our objective is to show that $f(p_i)$ is the Maxwellian 
distribution, with the normalization condition
\begin{equation}
\int_{-R}^Rf(p_i)dp_i = 1.
\label{eq-p_n}
\end{equation}

If the $i$th particle has coordinate $p_i$, the $(N-1)$ remaining particles 
share an energy less than the maximum energy $R^2-p_i^2$ on the $(N-1)$-sphere
\begin{equation}
p_1^2+p_2^2 \cdots +p_{i-1}^2 + p_{i+1}^2 \cdots +p_N^2 \leq R^2-p_i^2,
\label{eq-E1}
\end{equation} 
whose volume is $V_{N-1}(\sqrt{R^2-p_i^2})$. 
It can be easily proved that
\begin{equation}
V_N(R) = \!\int_{-R}^{R}\!V_{N-1}(\sqrt{R^2-p_i^2})dp_i.
\label{eq-theta1}
\end{equation}
Hence, the volume of the $N$-sphere for which the $i$th coordinate is
between $p_i$ and $p_i+dp_i$ is $V_{N-1}(\sqrt{R^2-p_i^2})dp_i$.
We normalize it to satisfy Eq.~(\ref{eq-p_n}), and obtain
\begin{equation}
f(p_i) = {V_{N-1}(\sqrt{R^2-p_i^2})\over V_N(R)},
\label{eq-f_n}
\end{equation}
whose final form, after some calculation is
\begin{equation}
f(p_i) = C_N R^{-1}\Big(1-{p_i^2\over R^2} \Big)^{N-1\over 2},
\label{eq-mm}
\end{equation}
with
\begin{equation}
C_N = {1\over\sqrt{\pi}}{\Gamma({N+2\over 2})\over \Gamma({N+1\over 2})}.
\label{eq-cn}
\end{equation}
For $N\gg 1$, Stirling's approximation can be applied to 
Eq.~(\ref{eq-cn}), leading to
\begin{equation}
\lim_{N\gg 1} C_N \simeq {1\over\sqrt{\pi}}\sqrt{N\over 2}.
\label{eq-cc}
\end{equation}
If we call $\epsilon$ the mean energy per particle, 
$E=R^2=N\epsilon$, then in the limit of large $N$ 
we have
\begin{equation}
\lim_{N\gg 1}\left(1-{p_i^2\over R^2}\right)^{N-1\over 2}
\simeq e^{-{p_i^2/2\epsilon}}.
\label{eq-ee}
\end{equation}
The factor $e^{-{p_i^2/2\epsilon}}$ is found 
when $N\gg 1$ but, even for small $N$, it can be a good approximation 
for particles with low energies.
After substituting Eqs.~(\ref{eq-cc})--(\ref{eq-ee})
into Eq.~(\ref{eq-mm}), we obtain the Maxwellian distribution in the asymptotic regime $N\rightarrow\infty$ 
(which also implies $E\rightarrow\infty$):
\begin{equation}
f(p)dp = \sqrt{1\over 2\pi\epsilon}\,e^{-{p^2/2\epsilon}}dp,
\label{eq-gauss}
\end{equation}
where the index $i$ has been removed because the distribution is the same for each particle, 
and thus the velocity distribution can be obtained by averaging
over all the particles. 

Depending on the physical situation the mean energy per particle $\epsilon$
takes different expressions. For a one-dimensional gas in thermal equilibrium
we can calculate the dependence of $\epsilon$ on the temperature, which, as
in the microcanonical ensemble, can be calculated by differentiating the entropy
with respect to the energy. The entropy can be written as $S=-kN\!\int_{-\infty} 
^{\infty} f(p)\ln f(p)\,dp$, where $f(p)$ is given by Eq.~(\ref{eq-gauss})
and $k$ is the Boltzmann constant. 
If we recall that $\epsilon=E/N$, we obtain
\begin{equation}
S(E)= {1\over 2}kN\ln\left({E\over N} \right) + {1\over 2}kN(\ln(2\pi)+1).
\label{eq-s2}
\end{equation}
The calculation of the temperature $T$ gives
\begin{equation}
T^{-1}= \left({\partial S\over \partial E} \right)_N = {kN\over 2E} = {k\over 2\epsilon}.
\end{equation}
Thus $\epsilon=kT/2$, consistent with the equipartition theorem. 
If $p^2$ is replaced by ${1\over 2}mv^2$, the Maxwellian 
distribution is a function of particle velocity, as it is usually given
in the literature:
\begin{equation}
g(v)dv = \sqrt{m\over 2\pi kT}\,e^{-{mv^2/2kT}}dv.
\end{equation}

This shows that the geometrical image of the volume-based statistical ensemble
allows us to recover the same result than that obtained 
from the microcanonical and canonical ensembles \cite{lopez2007-1,klages}.
Also, it confirms for this case the equivalence among all these ensembles 
in the thermodynamic limit.

\section*{Derivation of the Boltzmann-Gibbs distribution}

Here we start by assuming $N$ agents, each one with 
coordinate $x_i$, $i=1,\ldots,N$, 
with $x_i\geq 0$ representing the wealth or money of the agent $i$,
and a total available amount of money $E$:
\begin{equation}
x_1+x_2+\cdots +x_{N-1}+x_N \leq E.
\label{eq-e}
\end{equation} 
Under random evolution rules for the exchanging of money among agents \cite{yakovenko1},
let us suppose that this system evolves in the interior of the $N$-dimensional pyramid 
given by Eq. (\ref{eq-e}). The role of the heat reservoir, that in this model
supplies money instead of energy, could be played by the state or by the bank system 
in western societies.  
The formula for the volume $V_N(E)$ of an equilateral $N$-dimensional pyramid 
formed by $N+1$ vertices linked by $N$ perpendicular sides of length $E$ is
\begin{equation}
V_N(E) = {E^N\over N!}.
\label{eq-S_n1}
\end{equation}
We suppose that each point on the $N$-dimensional pyramid is equiprobable, 
then the probability $f(x_i)dx_i$ of finding 
the agent $i$ with money $x_i$ is proportional to the 
volume formed by all the points into the $(N-1)$-dimensional pyramid 
having the $i$th-coordinate equal to $x_i$. 
Our objective is to show that $f(x_i)$ is the Boltzmann factor
(or the Maxwell-Bolztamnn distribution), with the normalization condition
\begin{equation}
\int_{0}^Ef(x_i)dx_i = 1.
\label{eq-p_n1}
\end{equation}

If the $i$th agent has coordinate $x_i$, the $N-1$ remaining agents 
share, at most, the money $E-x_i$ on the $(N-1)$-dimensional pyramid
\begin{equation}
x_1+x_2 \cdots +x_{i-1} + x_{i+1} \cdots +x_N\leq E-x_i,
\label{eq-e1}
\end{equation} 
whose volume is $V_{N-1}(E-x_i)$. 
It can be easily proved that
\begin{equation}
V_N(E) = \!\int_{0}^{E}\!V_{N-1}(E-x_i) {dx_i }.
\label{eq-theta11}
\end{equation}

Hence, the volume of the $N$-dimensional pyramid for which the $i$th 
coordinate is between $x_i$ and $x_i+dx_i$ is $V_{N-1}(E-x_i)dx_i$.
We normalize it to satisfy Eq.~(\ref{eq-p_n1}), and obtain
\begin{equation}
f(x_i) = {V_{N-1}(E-x_i)\over V_N(E)},
\label{eq-f_n1}
\end{equation}
whose final form, after some calculation is
\begin{equation}
f(x_i) = NE^{-1}\Big(1-{x_i\over E} \Big)^{N-1},
\label{eq-mm1}
\end{equation}
If we call $\epsilon$ the mean wealth per agent, 
$E=N\epsilon$, then in the limit of large $N$ 
we have
\begin{equation}
\lim_{N\gg 1}\left(1-{x_i\over E}\right)^{N-1}
\simeq e^{-{x_i/\epsilon}}.
\label{eq-ee1}
\end{equation}
The Boltzmann factor $e^{-{x_i/\epsilon}}$ is found 
when $N\gg 1$ but, even for small $N$, it can be a good approximation 
for agents with low wealth. After substituting Eq.~(\ref{eq-ee1})
into Eq.~(\ref{eq-mm1}), we obtain the Maxwell-Boltzmann distribution 
in the asymptotic regime $N\rightarrow\infty$ (which also implies $E\rightarrow\infty$):
\begin{equation}
f(x)dx = {1\over \epsilon}\,e^{-{x/\epsilon}}dx,
\label{eq-gauss11}
\end{equation}
where the index $i$ has been removed because the distribution is the same for each agent, 
and thus the wealth distribution can be obtained by averaging over all the agents. 
This distribution has been found to fit the real distribution of incomes 
in western societies\cite{yakovenko1}.

Depending on the physical situation the mean wealth per agent $\epsilon$
takes different expressions and interpretations. 
For instance, doing a thermodynamic simile, 
we can calculate the dependence of $\epsilon$ on the temperature, which,
as in the microcanonical ensemble, can be obtained in this case by differentiating 
the entropy with respect to the total wealth. The entropy can be written as $S=-kN\!\int_{0} 
^{\infty} f(x)\ln f(x)\,dx$, where $f(x)$ is given by Eq.~(\ref{eq-gauss11})
and $k$ is the Boltzmann constant. 
If we recall that $\epsilon=E/N$, we obtain
\begin{equation}
S(E)= kN\ln\left({E\over N} \right) + kN.
\end{equation}
The calculation of the temperature $T$ gives
\begin{equation}
T^{-1}= \left({\partial S\over \partial E} \right)_N = {kN\over E} = {k\over \epsilon}.
\end{equation}
Thus $\epsilon=kT$, and the Boltzmann-Gibbs distribution 
is obtained as it is usually given in the literature:
\begin{equation}
f(x)dx = {1\over kT}\,e^{-x/kT}dx.
\end{equation}

This shows that the geometrical image of the volume-based statistical ensemble
allows us to recover the same result than that obtained 
from the microcanonical and canonical ensembles \cite{lopez2007-2}.
Also, it confirms for this case the equivalence among all these ensembles 
in the thermodynamic limit.

\section*{General derivation of the asymptotic distribution: an open problem}

Now the problem is stated in a general way.
Let $b$ be a real constant. 
If we have a set of positive variables $(x_1,x_2,\ldots,x_N)$ verifying 
\begin{equation}
x_1^b+x_2^b+\cdots +x_{N-1}^b+x_N^b \leq E
\label{eq-Ek}
\end{equation}
with an adequate mechanism assuring 
the equiprobability of all the possible states $(x_1,x_2,\ldots,x_N)$
into the volume given by expression (\ref{eq-Ek}),
will we have for the generic variable $x$ the distribution
\begin{equation}
f(x)dx \sim \epsilon^{-1/b}\,e^{-{x^b/b\epsilon}}dx,
\label{eq-gaussn}
\end{equation}
when we average over the ensemble in the limit $N\rightarrow\infty$?. 

Let us suppose that the answer to this last question is affirmative 
(as it will be probably shown in a next paper). If we define
\begin{equation}
c_b=\left[\!\int_{0} ^{\infty} e^{-y^b/b}\,dy\right]^{-1},
\label{eq-cb1}
\end{equation}
then expression (\ref{eq-gaussn}), redefined as    
\begin{equation}
f(x)dx = c_b \epsilon^{-1/b}\,e^{-{x^b/b\epsilon}}dx,
\label{eq-gaussn1}
\end{equation}
is normalized, i.e., $\!\int_{0} ^{\infty} f(x)dx=1$.
Following the thermodynamic simile done in the cases $b=1,2$, 
we can calculate the dependence of $\epsilon$ on the temperature 
by differentiating the entropy with respect to the energy. 
The entropy can be written as $S=-kN\!\int_{0} 
^{\infty} f(x)\ln f(x)\,dx$, where $f(x)$ is given by Eq.~(\ref{eq-gaussn1})
and $k$ is the Boltzmann constant. 
If we recall that $\epsilon=E/N$, we obtain
\begin{equation}
S(E)= {kN\over b}\ln\left({E\over N} \right) + {kN\over b}(1-b\ln c_b),
\end{equation}
where it has been used that $\epsilon=<x^b>=\!\int_{0} ^{\infty} x^bf(x)dx$.
Let us recall at this point that, for the case $b=2$, the limits used in 
the normalization integral of $f(x)$ in expression (\ref{eq-p_n}), and, 
therefore, in the calculation of $S(E)$,
are from $-\infty$ to $\infty$ instead from $0$ to $\infty$
that are used here. This does not introduce any change in the final result,
only redefines the constant $c_{b=2}$, which now is $\sqrt{2\over \pi}$
instead of the factor $1\over\sqrt{2\pi}$ from expression (\ref{eq-gauss}).

The calculation of the temperature $T$ gives
\begin{equation}
T^{-1}= \left({\partial S\over \partial E} \right)_N = {kN\over bE} = {k\over b\epsilon}.
\end{equation}
Thus $\epsilon=kT/b$, a result that recovers the theorem of equipartition of energy
for the quadratic case $b=2$.
The distribution for all $b$ is finally obtained:
\begin{equation}
f(x)dx = c_b\left({b\over kT}\right)^{1/b}\,e^{-x^b/kT}dx.
\end{equation}

This shows that the geometrical image of the volume-based statistical ensemble
allows us to recover the same result than that obtained 
from the microcanonical and canonical ensembles \cite{lopez2007-1,lopez2007-2}.
Also, it confirms for this case the equivalence among all these ensembles 
in the thermodynamic limit.

\section*{APPENDIX: \underline{A microcanonical image of the canonical ensemble}}

We are interested in this paper with alternative views of the different
statistical ensembles. Here we give a different image of the canonical ensemble
from that that is its usual presentation in the literature.

Let us suppose that a system with mean energy $\bar E$, and in thermal equilibrium
with a heat reservoir, is observed during a very long period $\tau$ of time.
Let $E_i$ be the energy of the system at time $i$. Then we have:
\begin{equation}
E_1+E_2+\cdots +E_{\tau-1}+E_{\tau} = \tau\cdot\bar E.
\label{eq-eee}
\end{equation} 
If we repeat this process of observation a huge number (toward infinity) of times,
the different vectors of measurements, $(E_1,E_2,\ldots,E_{\tau-1},E_{\tau})$,
with $0\leq E_i\leq \tau\cdot\bar E$, will finish by covering equiprobably the
whole surface of the $\tau$-dimensional hyperplane given by Eq. (\ref{eq-eee}).
If it is now taken the limit $\tau\rightarrow\infty$, the asymptotic probability
$p(E)$ of finding the system with an energy $E$ (where the index $i$ has been removed),
\begin{equation}
p(E)\; \sim \;\; e^{-E/\bar E},
\label{eq-e22}
\end{equation} 
is found by means of the geometrical arguments exposed in Ref. \onlinecite{lopez2007-2}.
Doing a thermodynamic simile, the temperature $T$ can also be calculated. It is
obtained that
\begin{equation}
\bar E = kT.
\label{eq-e222}
\end{equation} 
The {\it stamp} of the canonical ensemble, namely, the Boltzmann factor,
\begin{equation}
p(E)\;\sim\;\; e^{-E/kT},
\label{eq-e223}
\end{equation} 
is finally recovered from this new image of the canonical ensemble.

%\newpage

\end{document}